\documentclass[10pt]{article}
\usepackage{epsfig,graphicx}
\usepackage{times,ch2005}

\def\beq{\begin{equation}}
\def\eeq#1{\label{#1}\end{equation}}
\def\eeqn{\end{equation}}

\def\beqa{\begin{eqnarray}}
\def\eeqa#1{\label{#1}\end{eqnarray}}
\def\eeqan{\end{eqnarray}}

\def\Dslash{\not{\hbox{\kern-4pt $D$}}}
\def\dslash{\not{\hbox{\kern-2pt $\del$}}}

\newcommand{\tev}{\ensuremath{\mathrm{\,Te\kern -0.1em V}}\xspace}
\newcommand{\gev}{\ensuremath{\mathrm{\,Ge\kern -0.1em V}}\xspace}
\newcommand{\mev}{\ensuremath{\mathrm{\,Me\kern -0.1em V}}\xspace}
\newcommand{\kev}{\ensuremath{\mathrm{\,ke\kern -0.1em V}}\xspace}
\newcommand{\ev}{\ensuremath{\mathrm{\,e\kern -0.1em V}}\xspace}
\newcommand{\gevc}{\ensuremath{{\mathrm{\,Ge\kern -0.1em V\!/}c}}\xspace}
\newcommand{\mevc}{\ensuremath{{\mathrm{\,Me\kern -0.1em V\!/}c}}\xspace}
\newcommand{\gevcc}{\ensuremath{{\mathrm{\,Ge\kern -0.1em V\!/}c^2}}\xspace}
\newcommand{\mevcc}{\ensuremath{{\mathrm{\,Me\kern -0.1em V\!/}c^2}}\xspace}

\def\mus  {\ensuremath{\rm \,\mus}\xspace}

\def\mus        {\ensuremath{\,\mu{\rm s}}\xspace}    

\newcommand{\spc}{Smart Pixel Camera }
\newcommand{\hess}{H.E.S.S. }

\begin{document}


\Title{A Smart Pixel Camera for future Cherenkov Telescopes}
\bigskip


%
\label{HermannStart}

%
\author{ G.~Hermann, S.~Carrigan, B.~Gl\"uck, D.~Hauser\index{Hermann, G.} }

%
\address{Max-Planck-Institut f\"ur Kernphysik\\
Saupfercheckweg 1 \\
D-69115 Heidelberg, Germany \\
}

\makeauthor\abstracts{
The \spc is a new camera for imaging atmospheric Cherenkov telescopes,
suited for a next generation of large multi-telescope
ground based gamma-ray observatories. The design of the camera foresees
all electronics needed to process the images to be located inside the 
camera body at the focal plane.
The camera has a modular
design and is scalable in the number of pixels.
The camera electronics provides the performance needed for the 
next generation instruments, like short
signal integration time, topological trigger and short trigger gate,
and at the same time the design 
is optimized to minimize the cost per channel. 
In addition new features are implemented, like the measurement 
of the arrival time of light pulses in the pixels on 
the few hundred psec timescale.
The buffered readout system of the camera allows to take images at sustained
rates of O(10 kHz) with a dead-time of only about 0.8 \% per kHz.
}

\section{Introduction}

Ground-based gamma-ray astronomy has achieved a major breakthrough with the 
new results from the \hess experiment, which has reached a critical 
sensitivity, and allows for detailed morphological, spectroscopic and temporal 
investigations of a large number of VHE gamma-ray sources. 
Future projects beyond \hess will aim for several
goals: lowering the energy threshold into the 5-10 GeV energy range,
improving the sensitivity in the energy
range up to about 100 TeV 
by an order of magnitude, and improving 
the survey capability of the instruments. 
Ideas for such projects have already been described before
with emphasis on lowering the energy threshold \cite{felix_5at5}. 
Systems of imaging atmospheric Cherenkov telescopes (IACT) at high altitude, 
with O(100) telescopes are amongst the new concepts. The cameras will 
have a field of view about 
5$^\circ$-10$^\circ$, covered by several thousand pixels, and have to be 
able to record images at rates up to several kHz. As for the present generation
of cameras, a short signal integration time in the 10-20 nsec range and 
very short coincidence trigger gates are mandatory. Due to the large number
of telescopes, the cost for the cameras will be one 
of the critical parameters in the layout of such systems. Therefore a 
camera design is needed that provides the required performance and in 
parallel minimizes the cost per channel. In addition maintainability and 
stability of operation are important design considerations. Features
that allow to continuously monitor the performance of the camera
are needed and an efficient commissioning is critical in systems 
with O(100) telescopes.\\
The concept of the Smart Pixel was presented for the first time in 1999 
\cite{spc_snowbird} in the context of the \hess experiment. 
However, this development was not pursued at the time;  it was 
resumed with significant modifications by the end of 2002. Below we
describe the design of the prototype of a \spc and focus
on the {\it camera electronics} that has been developed at the MPI f\"ur Kernphysik
in Heidelberg. We do not describe the photon detectors (photomultipliers) and
the HV supplies, since the design of the \spc does not depend specifically on
them. The devices used in the prototype are identical to the ones 
in the current \hess cameras (Photonis XP2960 and ISEG PHQ2960, 
respectively).  \\
In the following we first introduce the general architecture of 
the Smart Pixel Camera, then describe the pixels and their readout in  more detail
and finally show some first results from test measurements with a prototype.

\section{The Smart Pixel Camera}

\subsection{Architecture of  the Smart Pixel Camera}

\begin{figure}[tb]
\begin{center}
\epsfig{file=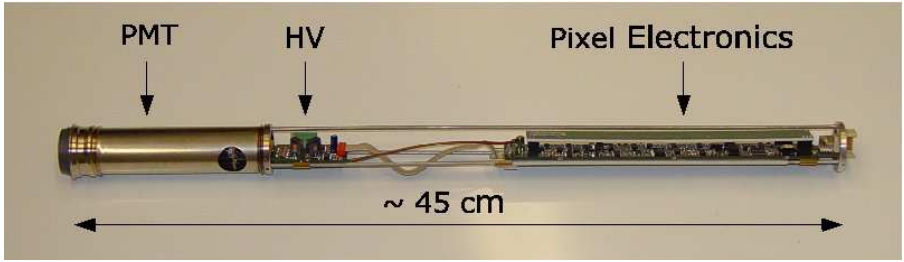,width=4.7in}
\caption{Photo of a Smart Pixel.}
\label{fig:pixel}
\end{center}
\end{figure}

  A schematic overview of the architecture of the  \spc is shown in
\begin{figure}[htb]
\begin{center}
\epsfig{file=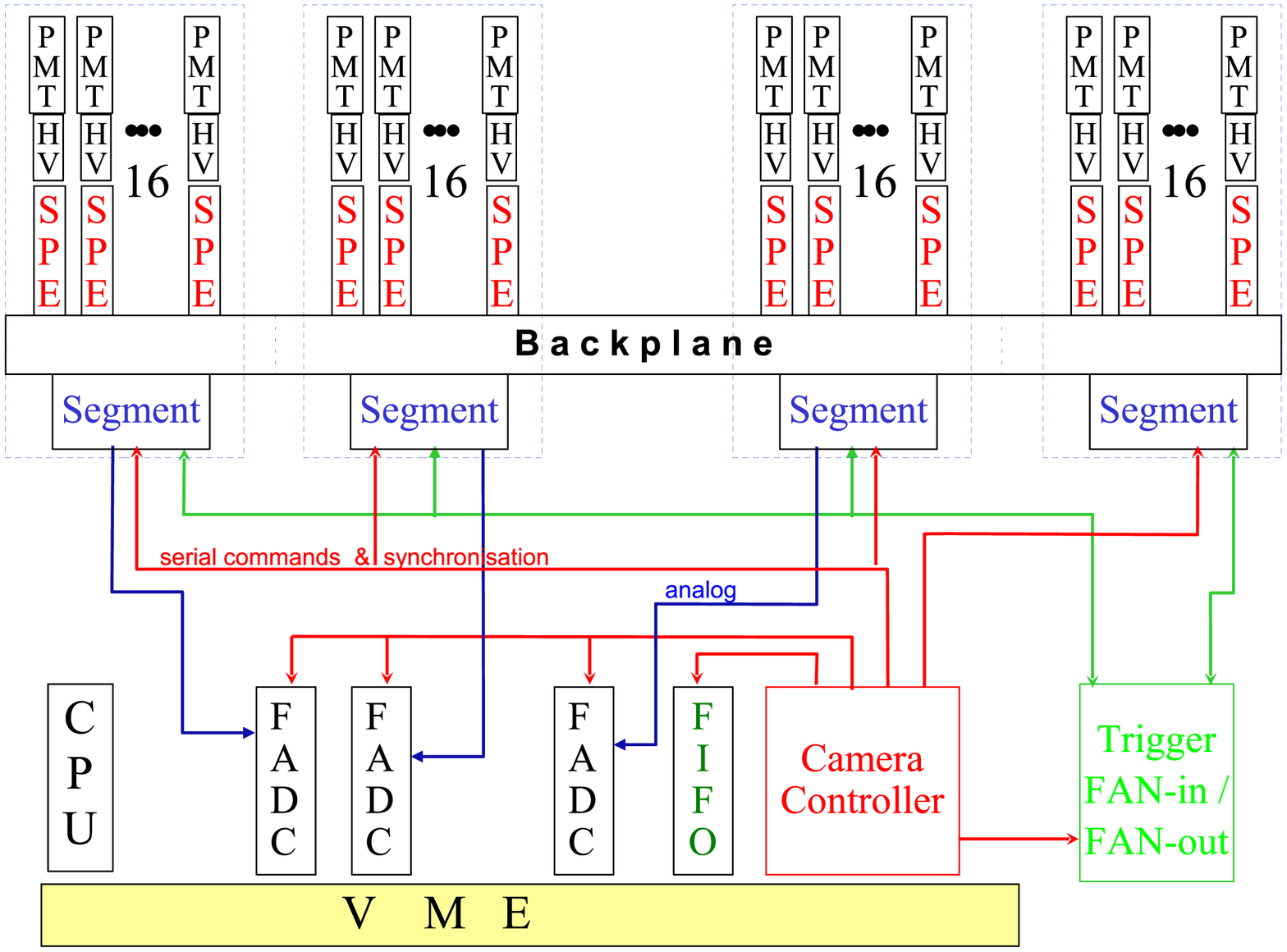,width=4.in}
\caption{Overview of the architecture of the \spc.}
\label{fig:spc_overview}
\end{center}
\end{figure}
  Figure \ref{fig:spc_overview}. The main building block of the camera
  is the Smart Pixel (Figure \ref{fig:pixel}), which consists of a 
  photomultiplier (PMT), 
  an active high voltage supply (HV) and the smart pixel electronics (SPE).
  The SPE contains all analog processing of the PMT 
  signal, including signal integration, timing measurement, monitoring and
  analog storage. All information coming from the pixel is provided as 
  analog levels, which are multiplexed onto one analog line from each pixel.
  The trigger of the camera is also contained in the pixels. It consists
  of a programmable N out of 7 cluster coincidence, where each pixel is the
  center of a cluster of 7 pixels in the hexagonal pixel matrix of
  the camera. The pixels are plugged into a segmented backplane, where
  each segment serves up to 16 pixels. The backplane provides lines for
  the distribution
  of the trigger signals, for the analog bus for the pixel signals,
  and for the  serial command bus to the pixels. 
  All functions on the pixel are controlled by the 
  {\it camera controller}  via the serial command bus 
  (Fig. \ref{fig:spc_overview}).   
  The power distribution is also done through the backplane. 
  Each segment is controlled by a {\it segment
  controller}, which has 3 main tasks: the distribution of 
  the digital commands to the pixels, coming from the camera controller, the
  multiplexing of the analog signals from the pixels to an ADC and to ``OR'' 
  the  trigger signals coming from the pixels and to distribute  them through
  the camera {\it trigger FAN-in/FAN-out}. \\
  The digitization of the analog output levels provided by the pixels
  is done using a FADC system running in
  multiplexer mode. The synchronization of the switching of the
  multiplexers on the pixels and on the segments and the digitization on the
  FADCs  are done by the {\it camera controller}. 
  The digitized signals are stored in multi-event buffers on the FADCs
  and can be readout asynchronously. This allows to decouple the front-end
  readout (digitization) of the camera from the back-end readout (reading 
  data into the CPU memory). Currently, for the readout of the FADCs and for 
  the programming of the pixels through the camera controller a VME-based 
  system is implemented. 
 \\
  Since all electronics of the camera is installed inside the focal plane box,
  the only interfaces are a power cable, the computer network connection 
  and a pair of optical fibers to connect the
  trigger of the camera to a multi-telescope trigger in case of 
  stereoscopic operation \cite{hess_trigger_paper}. 
  Therefore, the complete camera, including the trigger, 
  and the digitization and 
  readout system, can be tested and characterized in the lab 
  before installation into a telescope at some remote location.

 \subsection{The Smart Pixel Electronics}
    The analog signal from the PMT is split 
    into three paths: a path for signal integration, a trigger path and 
    a path to derive monitoring information.
  \subsubsection{The Pixel Trigger}
    In the trigger path the analog signal is fed into a high bandwidth 
    discriminator with programmable threshold and a minimal gate width 
    of about 4 nsec. The discriminator signal is then delayed by a 
    programmable delay 
    of 0 to 5 nsec in steps of 1~nsec, which is used to compensate for the different PMT 
    transit times, depending on the high voltage applied to the PMT. The 
    discriminator signal is then sent via the backplane to the six 
    neighboring pixels and also to an internal programmable logic 
    device (PLD), where the discriminator
    signals from the six neighboring pixel arrive. On the PLD a programmable 
    multiplicity of at least N pixels is required to release a trigger.
    This condition can only be fulfilled, when at least N neighboring
    pixels have generated a level 0 trigger within the coincidence time
    determined by the discriminator gate width.
    If the trigger condition is fulfilled, the pixel sends the
    trigger signal via the backplane to the segment controller, where the 
    trigger signals from all the pixels of a segment are "ORed"  and sent to
    a camera trigger FAN-in/FAN-out, which receives the trigger signals
    from all segments and distributes them simultaneously via the segment
    controllers to all pixels.

    In order to monitor the performance of the pixel trigger, the rate of 
    discriminator signals and the rate of triggers generated by the pixel
    are also provided as analog levels.

  \subsubsection{Signal Integration}
     In the signal integration path, the signal is first amplified and then
     delayed by about 90 nsec using a delay line, which is based on 
     a 12 layer printed
     circuit board (PCB). The PCB contains a structure of a meandering 
     electrical line embedded in ground lines and between 
     ground layers, shaped
     in a way to imitate a quasi-coaxial structure. The resulting pulse 
     after the ~90 nsec delay has a acceptable width of about 8 nsec FWHM. \\
     After the delay line, the signal is split into two paths with 
     a difference in amplification by a factor of
     18 (hi-gain and lo-gain path). 
     A V/I converter and a gated integrator are used to integrate the 
     charge of the two channels.
      The integration is started when a camera trigger occurs. The width of the
     integration gate is programmable in steps of 3 nsec from 10 nsec to 25
     nsec. Also the delay of the gate is programmable in steps of 1 nsec 
     from 0 to 5 nsec, again to compensate for individual transit time
     differences of the PMTs. The signals from the gated integrators are
     connected to the pixel multiplexer.

  \subsubsection{Pixel timing}
      The precise knowledge of the time structure of the Cherenkov images might
      help to suppress background events offline. 
      A cost-effective possibility is to use time to amplitude converters (TAC),
      which are implemented on the pixels using a constant current source and 
      a gated integrator. Upon a camera trigger, the TACs of
      the pixels are started. The discriminator signals are delayed internally 
      on the pixels by about 95
      nsec (a few nsec longer than it takes to generate and distribute the camera
      trigger) and used to stop the TAC. 
      The TACs therefore measure the relative trigger time of the pixel 
      discriminators with respect to the camera trigger.
      If a pixel has
      not triggered in an event, its TAC will run into an overflow.
      Besides the timing measurement, this information in addition  
      tells which pixels have triggered in a given event. The resolution of the
      time information is well below 1 nsec, depending on the pixel amplitude
      (see below). The trigger information can be used to precisely determine
      the trigger behavior of the discriminator (see below).

 \subsection{Digitization scheme and dead time}
       The pixels provide their event information (hi-gain and lo-gain 
       amplitude and time information)
       and monitoring information, like PMT anode current, 
       pixel trigger rate and temperatures as analog levels.
       As explained above, the pixel information is multiplexed onto one 
       analog line. The analog lines of 16 pixels are then multiplexed onto
       one analog line for the segment, which is  connected to a FADC
       channel. By switching the segment and pixel multiplexers, 
       the corresponding signals can be selected and digitized.
       For "normal" events, the two amplitudes (hi-gain and lo-gain) and 
       the time information per pixel are digitized. In monitoring events,
       also the remaining values are digitized.
       The switching of the multiplexers and the digitization is done at
       a frequency of 10 MHz.
       Since each segment with 16 pixels is connected to one FADC channel, 
       the digitization of an event with 3 event-related values per pixel
       takes only 4.8~$\mu$sec. Since the data is written directly
       into a multi-event buffer on the FADC, the front-end dead time per event 
       amounts to only 4.8~$\mu$sec plus about 2 times 1.5~$\mu$sec charge-up and reset
       time of the analog levels on the pixel. This dead time is independent
       of the number of pixels or segments in the camera, since the digitization
       of different segments is done in parallel. \\
\begin{figure}[tb]
\begin{center}
\epsfig{file=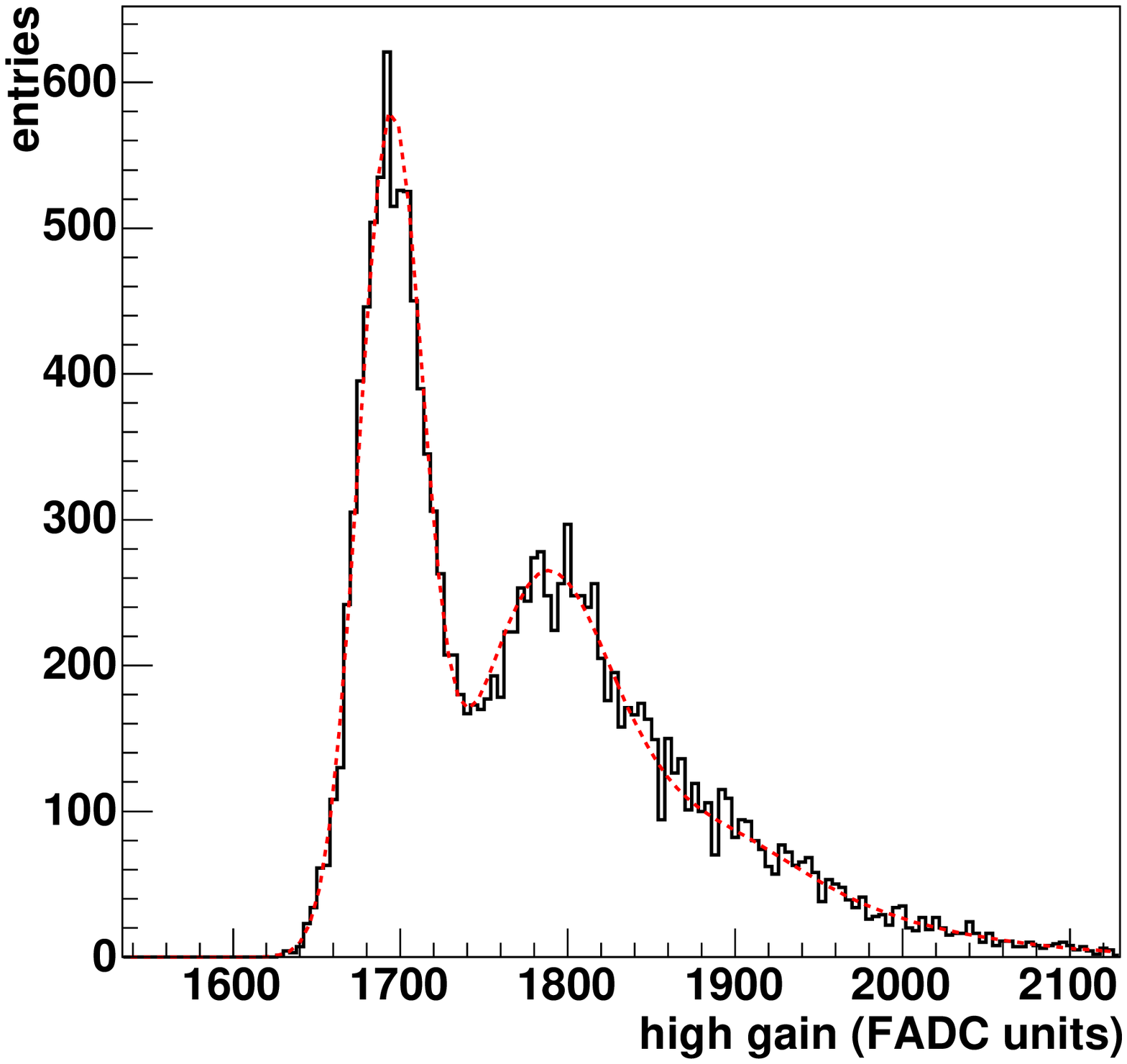,height=2.0in}
\epsfig{file=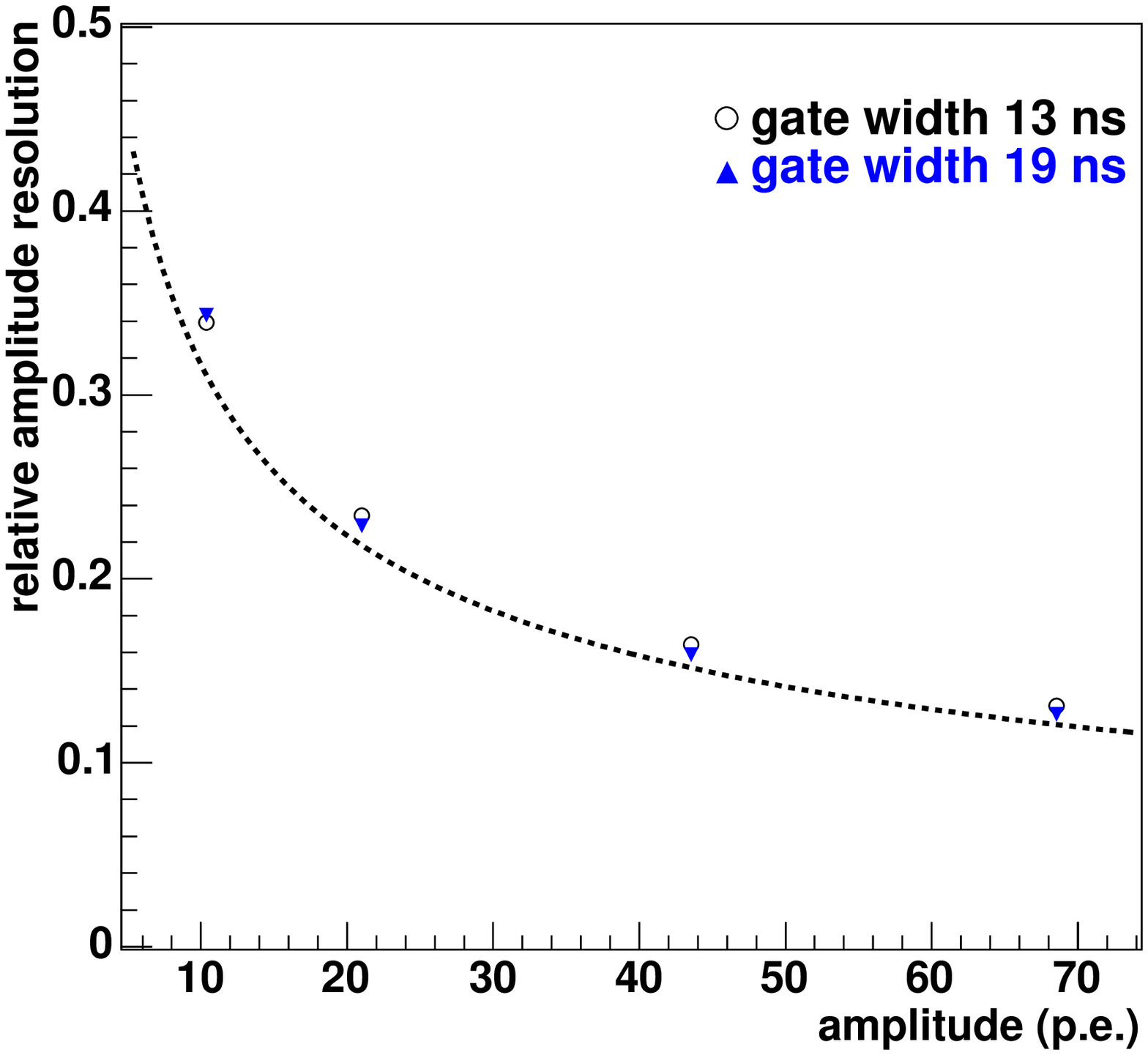,height=2.0in}
\caption{a) Single photoelectron spectrum of a pixel. b) 
 Relative amplitude resolution of the pixels as a function of the light pulse
 ampitude. The line describes the 1/$\sqrt{\mbox{N}_{p.e.}}$ law, which is the
 limit given by poisson fluctuations of the number of 
 photoelectrons at the photocathode.}
\label{fig:spe_resolution}
\end{center}
\end{figure}
       The event buffers on the FADCs (Model: SIS3300) can be readout 
       asynchronously with 64 Bit block-transfer in DMA mode via the VME bus,
       while new events are
       written to the buffers. With this data transfer scheme a rate of about 
       38 MByte/sec has been reached. Due to the decoupling of the front-end 
       from the readout of the back-end, up to a mean event rate of about 5-6 kHz, 
       the dead time of the camera is given only by the front-end dead time of
       about 8~$\mu$sec / event. If the camera uses significantly 
       more than about 1000 pixel or operates at a higher mean event rate, the 
       readout of the FADCs has to be done using more than one CPU 
       (on a segmented VME backplane) and the
       back-end bandwidth of the system then scales with the number of CPUs used.

\begin{figure}[htb]
\begin{center}
\epsfig{file=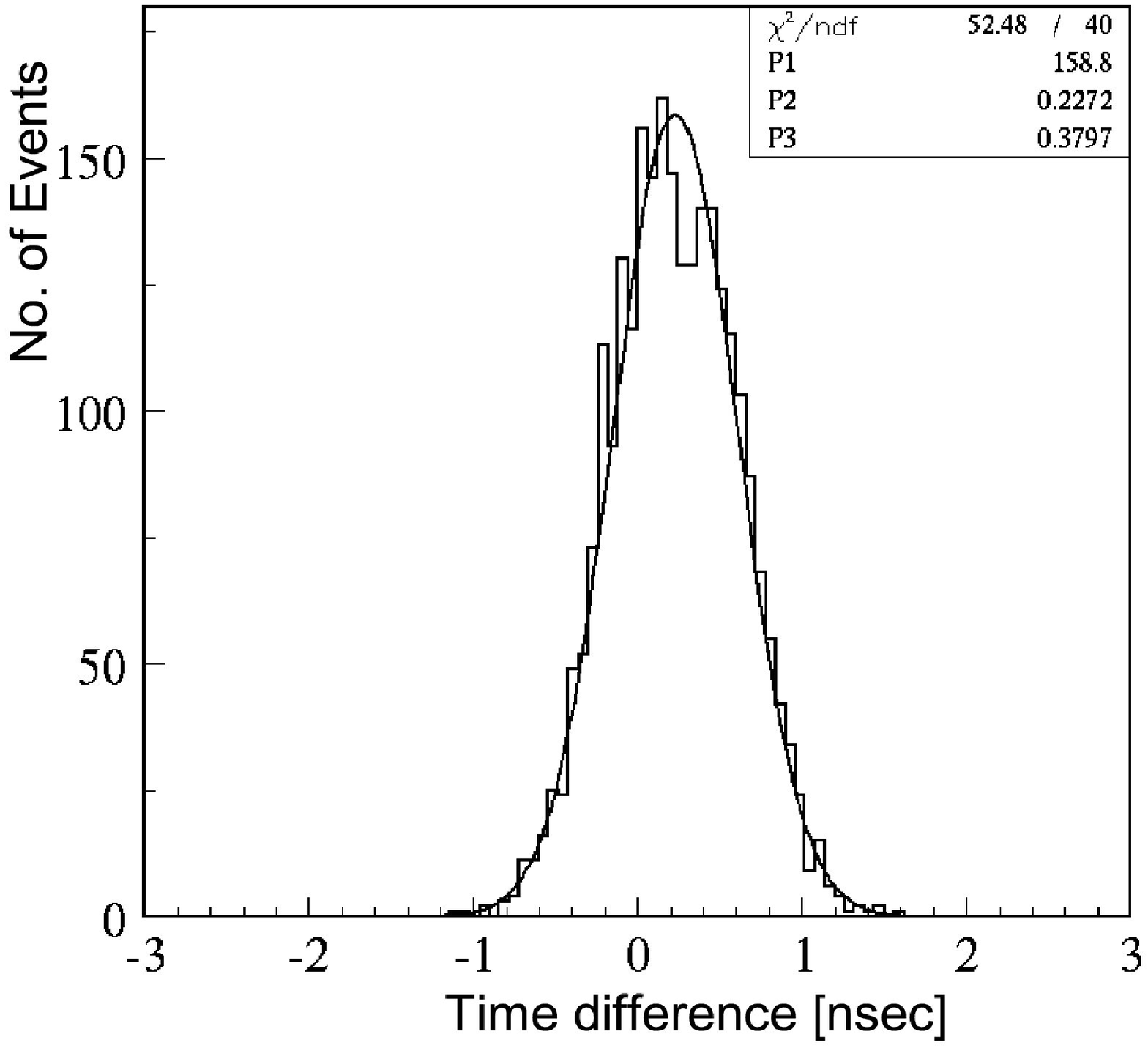,height=2.0in}
\epsfig{file=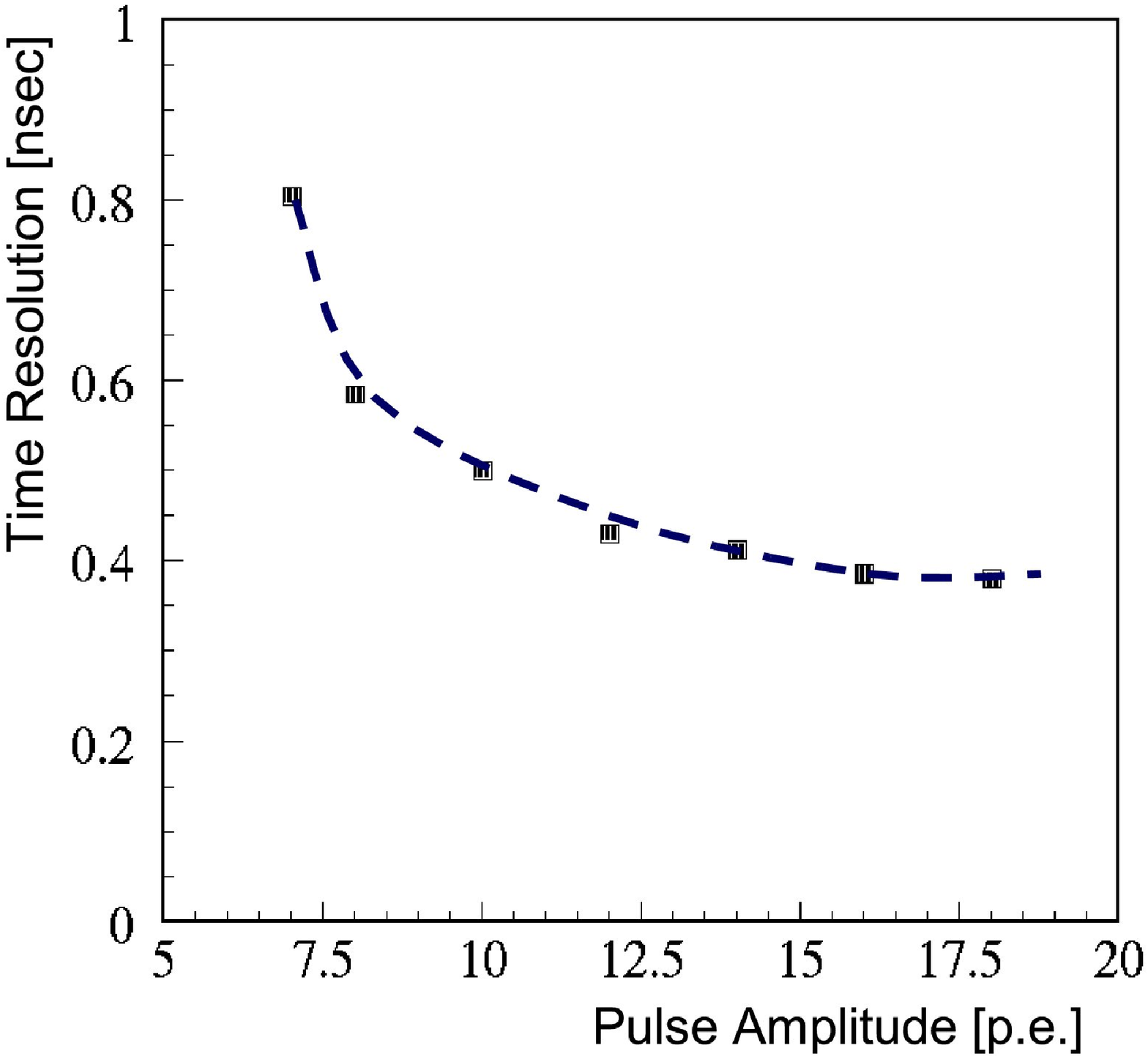,height=2.0in}
\caption{a) Histogram of the time difference of the trigger 
signals of two pixels. b)
Time resolution of a pixel as a function of the pulse amplitude, 
when illuminated
with a pulsed LED with 3 nsec FWHM pulse width.}
\label{fig:tac_resolution}
\end{center}
\end{figure}

\begin{figure}[htb]
\begin{center}
\epsfig{file=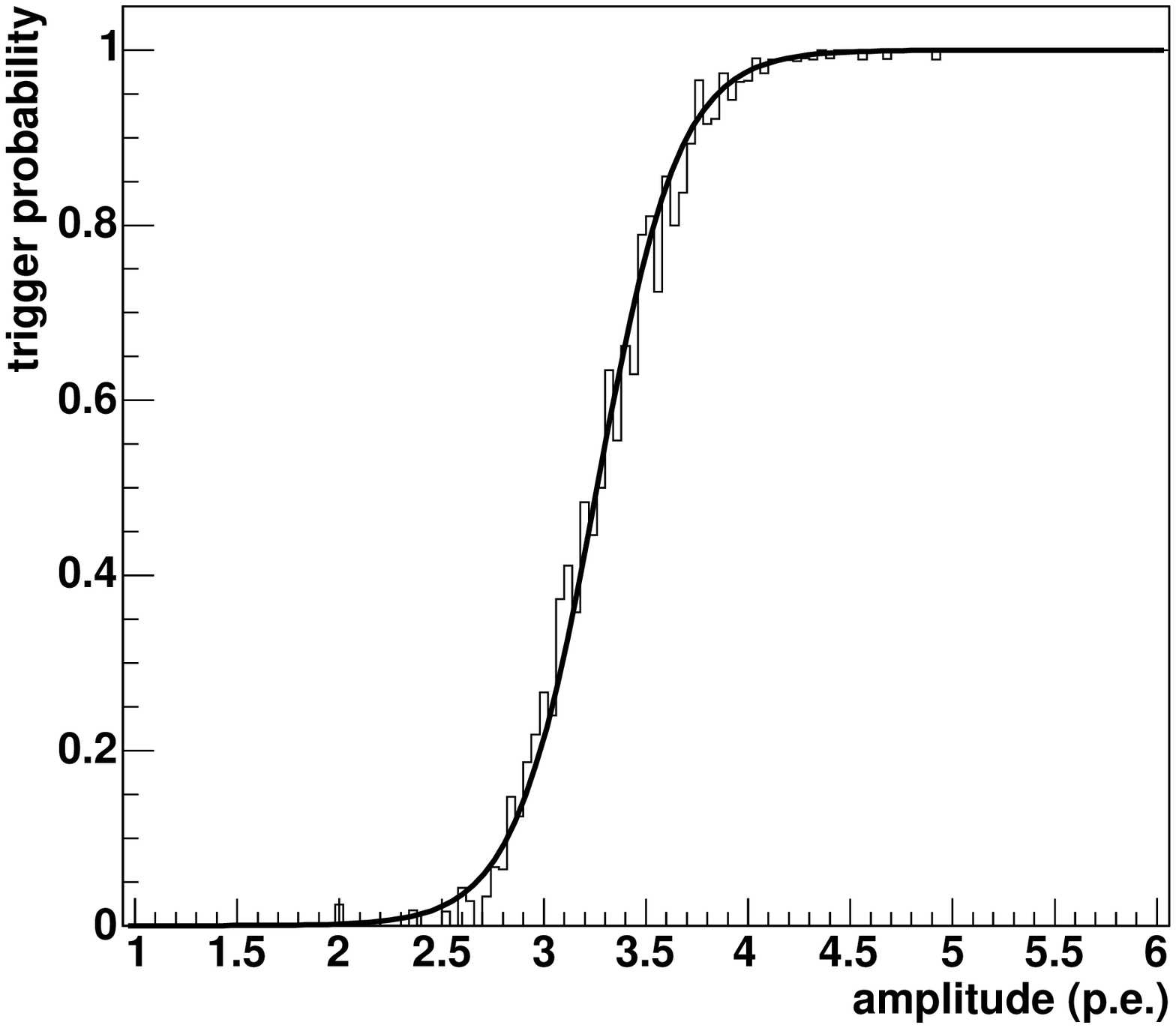,width=2.in}
\epsfig{file=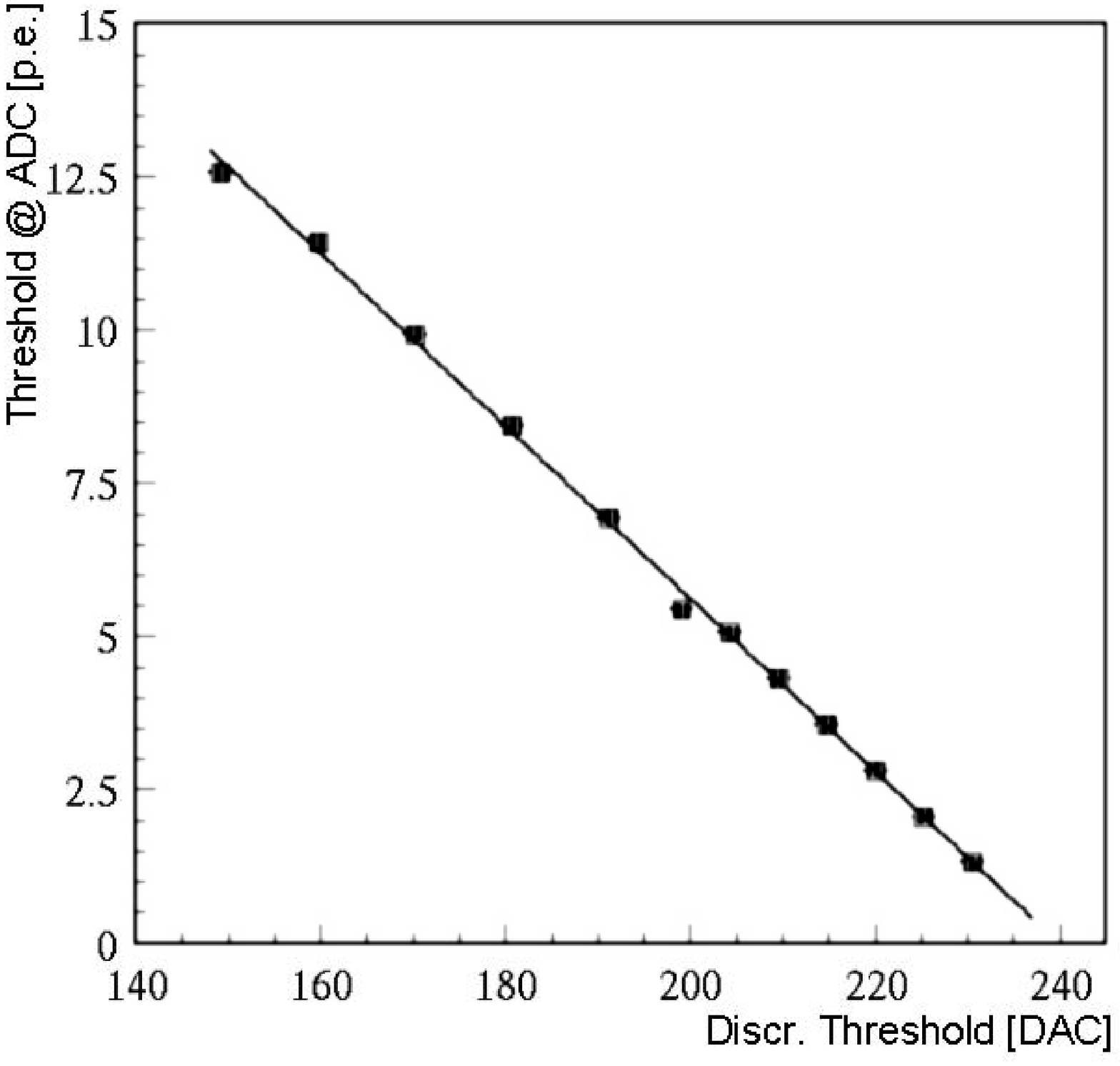,width=1.8in}
\caption{a) Discriminator threshold function of a pixel. b) Measured trigger 
threshold of a pixel as a function of the threshold applied at the
discriminator.}
\label{fig:trg_thr}
\end{center}
\end{figure}

\section{First Test Measurements}
After two iterations of prototyping of the pixel electronics, a larger batch
of pixels has been produced in order to be able to test the performance
of a prototype camera after system integration. The prototype camera
contains 64 pixels (as of April 2005) and uses the full digitization and
triggering scheme, but also the final power supply and cooling system, in 
order to test possible impact of these devices on the performance or noise
of the system. The electronics is mounted in a camera housing, which can
contain up to 1020 pixels and which can be used in a future prototype 
telescope. The camera is set up in a dark room, where it can be illuminated
with a high rate of LED pulses with a pulse width of 3~nsec and amplitudes 
of a few photoelectrons per pixel, and also with a pulsed N{$_2$}-laser 
through a scintillator and a programmable neutral density filter wheel, with 
3~nsec pulses and intensities up to a few thousand photoelectrons per pixel.
Figure \ref{fig:spe_resolution} a) shows an example of a single photoelectron spectrum, 
(as obtained by illumination with the pulsed LED) together with a fit to the
spectrum. From the fit one obtains a typical value of 0.2~p.e. electronics noise
of the hi-gain channel and an amplitude resolution of 0.45~p.e. of the single
photoelectron peak. In figure \ref{fig:spe_resolution} b) the amplitude 
resolution of the hi-gain channels as a function of the light 
pulse amplitude is displayed. It can be seen that the resolution is
close to the limit given by poisson fluctuations of the number of 
photoelectrons at the photocathode.
The resolution of the TAC is demonstrated in figure \ref{fig:tac_resolution} a), 
where the time difference between two pixels is plotted. As expected, the 
time resolution depends strongly on the light amplitude (figure 
\ref{fig:tac_resolution} b)). At amplitudes close to the discriminator
threshold the resolution is about 1~nsec, and reduces to $<$ 500~psec at
amplitudes larger than 10~p.e. . \\ As mentioned above, the time information of
each pixel also tells if a pixel has triggered in a given event. Therefore
the trigger information of a pixel can be related to its amplitude information,
allowing to measure precisely the trigger threshold of each pixel during normal
operation. This is illustrated in figure \ref{fig:trg_thr} a), where the
trigger probability of a pixel is plotted against the measured light amplitude.
Figure \ref{fig:trg_thr} b) shows the measured trigger threshold of a pixel
in units of photoelectrons as a function of the preset threshold in digital
units. Such measurements can be used to calibrate and monitor the
pixel thresholds.

\section{Summary and Outlook}
The \spc is a promising concept for future IACT systems. The first
tests indicate that the camera provides the performance needed
in such telescope systems. Due to the highly multiplexed readout
and the high integration of the electronics, the total cost per 
channel of such a camera is significantly below
current designs. Further tests with up to 200 pixels are planned.

\section*{Acknowledgments}
We would like to thank the staff of the electronics and mechanical workshops at
the Max-Planck-Institut f\"ur Kernphysik in Heidelberg, who have
been working with great enthusiasm on the development of the \spc.

%
\label{HermannEnd}
 
\end{document}